\documentclass[aps,prl,twocolumn,superscriptaddress,floatfix]{revtex4}

\usepackage{graphicx}
\usepackage{amsmath}
\usepackage{amssymb}
\usepackage[amssymb]{SIunits}
\usepackage{bm}

\newcommand{\e}{\mathrm{e}}
\newcommand{\kB}{\mathrm{k_B}}
\newcommand{\lB}{\ell_{\mathrm{B}}}

\newcommand{\ave}[1]{\left \langle #1 \right \rangle}
\renewcommand{\vec}[1]{\bm{#1}}

\begin{document}

\title{ Salt-induced counterion-mobility anomaly in polyelectrolyte
  electrophoresis}

\author{Sebastian Fischer} \affiliation{Physik Department, Technische
  Universit{\"a}t M{\"u}nchen, 85748 Garching, Germany} %
\author{Ali Naji} \affiliation{Physik Department, Technische
  Universit{\"a}t M{\"u}nchen, 85748 Garching, Germany}
\affiliation{Materials Research Laboratory, \& Department of Chemistry
  and Biochemistry, University of California, Santa Barbara, CA 93106,
  USA} %
\author{Roland R.\ Netz} \affiliation{Physik Department, Technische
  Universit{\"a}t M{\"u}nchen, 85748 Garching, Germany}


\begin{abstract}
  We study the electrokinetics of a single polyelectrolyte chain in
  salt solution using hydrodynamic simulations.  The salt-dependent
  chain mobility compares well with experimental DNA data. The
  mobility of condensed counterions exhibits a salt-dependent change
  of sign, an anomaly that is also reflected in the counterion excess
  conductivity.  Using Green's function techniques this anomaly is
  explained by electrostatic screening of the hydrodynamic
  interactions between chain and counterions.
\end{abstract}

\pacs{}

\maketitle

Polyelectrolytes (PEs) are macromolecules with ionizable groups that
dissociate in aqueous solution and thus give rise to a charged PE
backbone and a diffusely bound cloud of neutralizing
counterions~\cite{barrat-joanny:1996}.  Numerous applications in
chemical, biological, and medical engineering rely on the response of
PEs to externally applied electric fields (E-fields), determined by a
balance of electrostatic and hydrodynamic effects and controlled by
various factors such as salt concentration, PE charge density,
etc.~\cite{viovy:2000}.  The simplest scenario providing a basic
testing ground for our understanding of PE dynamics in the dilute
limit is \emph{free-solution electrophoresis}, where a single PE chain
is subject to a homogeneous static
E-field~\cite{hartford-flygare:1975,%
  hoagland-arvanitidou-welch:1999,stellwagen-stellwagen:2003}.

Previous theoretical approaches combined mean-field electrostatics
with low Reynolds number hydrodynamics.  Solutions of the
electrokinetic equations were obtained
numerically~\cite{schellman-stigter:1977} or analytically using
counterion-condensation theory~\cite{manning:1981} and account for the
experimentally measured salt dependent electrophoretic mobilities of
biopolymers such as DNA or synthetic PEs.  Counterions in the
immediate vicinity of the PE chain were assumed to stick to and move
along with the PE under the action of the applied E-field.  This
assumption becomes crucial for the conductivity of PE solutions, and
indeed inconsistencies between experimental mobility and conductivity
studies are documented in literature, pointing to some basic riddles
in the coupling of PE and counterion dynamics in
E-fields~\cite{stigter:1979b}.  Pioneering explicit-water
all-atomistic simulations of PEs in E-fields have been
performed~\cite{yeh-hummer:2004}. Due to the immense computational
demand they are restricted to elevated field strengths, short PEs, and
short simulation times.  Implicit-solvent simulations have quite
recently addressed the molecular-weight-dependent PE mobility in the
salt-free case~\cite{grass-etal:2008,frank-winkler:2008} and yielded
good agreement with experiments.

In the present paper we use coarse-grained implicit-solvent
hydrodynamic simulations~\cite{ermak-mccammon:1978} and study the
salt-dependent electrophoretic response of a single PE.  By
replicating the PE periodically we eliminate finite-chain-length
effects.  We concentrate on the salt-dependent interplay of PE versus
counterion mobility in the infinite chain limit and show that the
condensed counterion mobility changes sign as a function of salt
concentration. For low salt, counterions stick to the PE and move
along in the E-field in agreement with the canonic viewpoint.  For
high salt, on the other hand, the motion decouples and counterions
move opposite to the PE.  This anomaly is captured by an analytic
theory developed here for weakly charged chains based on the
electrostatically screened hydrodynamic interaction tensor. For DNA
our simulations reproduce experimental salt-dependent mobilities
without fitting parameters and predict an experimentally measurable
anomaly of the counterion excess conductivity.  The counterion anomaly
is also directly accessible by NMR
experiments~\cite{boehme-scheler:2004} or PE conductivity studies in
nanopores or nanochannels~\cite{smeets-etal:2006}.

\begin{figure}[b!]
  \includegraphics[height=4.4cm]{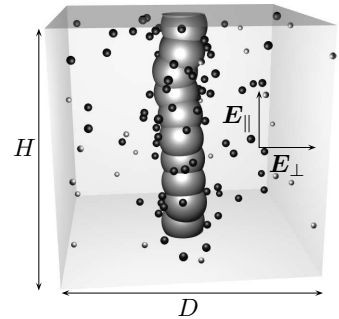}
  \caption{%
    Simulation cell for a DNA segment with counterions (dark grey) and
    coions (light grey).  Periodic boundary conditions are applied
    along the projected end-to-end distance $H$ of the DNA segment.
    The external electric field is applied either parallel ($\vec
    E_{||}$) or perpendicular ($\vec E_{\perp}$) to the PE axis.}
  \label{fig:1}
\end{figure}

\begin{figure*}
  \includegraphics{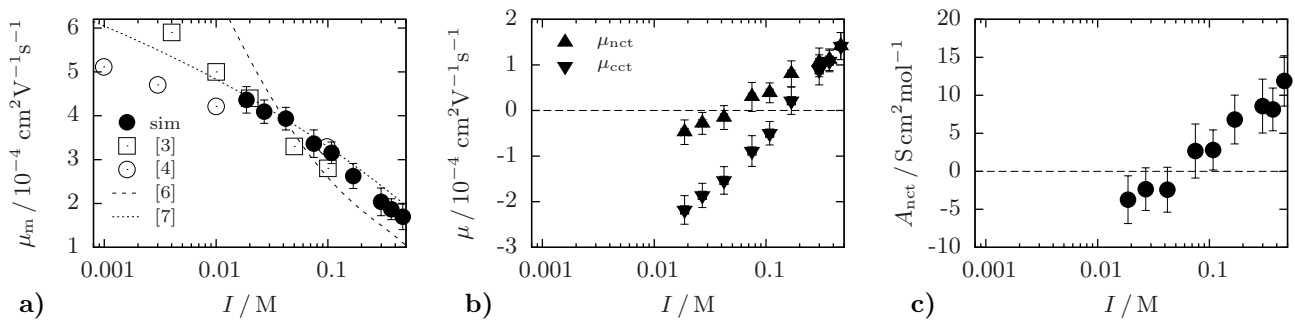}
  \caption{%
    Hydrodynamic simulations of DNA in aqueous NaCl solution of
    various ionic strengths $I$ at $\unit{20}{\celsius}$. %
    a) Electrophoretic DNA mobility $\mu_{\mathrm{m}}$ as obtained
    from simulations (filled symbols) and experiments (open symbols)
    as a function of $I$, compared to theories by
    Stigter~\cite{schellman-stigter:1977} (dashed line) and
    Manning~\cite{manning:1981} (dotted line). %
    b) Mobility of neutralizing counterions $\mu_{\mathrm{nct}}$, and
    condensed counterions $\mu_{\mathrm{cct}}$. %
    c) Counterion excess conductivity $A_{\mathrm{nct}}$ according to
    Eq.~(\ref{eq:anct}).}
  \label{fig:2}
\end{figure*}

In our hydrodynamic simulations we consider a PE consisting of charged
beads together with neutralizing counterions and added symmetric salt,
cf.\ Fig.~\ref{fig:1}.  The vertical box height $H$ and lateral width
$D$ are fluctuating while keeping the volume $HD^2$ and thus the
concentration of monomers $c_{\mathrm{m}}$, neutralizing counterions
$c_{\mathrm{nct}}$ and salt ion pairs $c_{\mathrm{s}}$ fixed.
Periodic boundary conditions along the vertical axis are implemented
by coupling the box height $H$ to the vertical PE extension.  All
particle positions $\vec r_i$ evolve according to the position
Langevin equation, $ \dot{ \vec r}_i(t) = - \sum_j \vec{\mathrm
  M}_{ij} \cdot \nabla_{\vec r_j} U(t) + \vec \xi_i (t) $.  The
thermal coupling is modeled by a Gaussian white noise with $\ave{ \vec
  \xi_i(t)} = \vec 0$ and $\ave{ \vec \xi_i (t) \vec \xi_j (t')} = 2
\kB T \, \vec{\mathrm M}_{ij} \delta (t-t')$ according to the
fluctuation-dissipation theorem.  Hydrodynamic interactions are
included via the Rotne-Prager-Yamakawa mobility tensor $\vec{\mathrm
  M}_{ij}$~\cite{ermak-mccammon:1978}, which accounts for finite
hydrodynamic particle radii $a_i$ ($a_i = a_{\mathrm{m}},
a_{\mathrm{ct}}, a_{\mathrm{co}}$ for monomers, counterions, and
coions).  The interaction potential $U = U_{\mathrm{LJ}} +
U_{\mathrm{C}} + U_{\mathrm{S}} + U_{\mathrm{ext}}$ consists of: i) A
truncated, shifted Lennard-Jones potential, $ U_{\mathrm{LJ}} / \kB T
= \epsilon \sum_{\langle ij \rangle} \left[ ( \sigma_{ij} / r_{ij}
  )^{12} -2 ( \sigma_{ij} / r_{ij} )^{6} +1 \right] $ for $r_{ij} \leq
\sigma_{ij} $ between ions and monomers that prevents electrostatic
collapse of opposite charges, where $r_{ij} = |\vec r_i - \vec r_j|$
is the distance between particles $i$ and $j$ and $\sigma_{ij}= (a_{i}
+ a_{j}) / 2$ and $\epsilon$ define the soft-core distance and
repulsion strength.  ii) An unscreened Coulomb potential $
U_{\mathrm{C}} / \kB T = \lB \sum_{\langle ij\rangle} q_i q_j / r_{ij}
$, where $q_i$ denotes particle valency ($q_i = q_{\mathrm{m}},
q_{\mathrm{ct}}, q_{\mathrm{co}}$ for monomers, counterions, and
coions) and $\lB = \e^2 / 4 \pi \epsilon_{\mathrm{r}} \epsilon_0 \kB
T$ is the Bjerrum distance at which two unit charges interact with
thermal energy $\kB T$ ($\lB = \unit{7.21}{\angstrom}$ in water at
$\unit{20}{\celsius}$).  iii) A harmonic potential, $ U_{\mathrm {S}}
/ \kB T = (K / 2) \sum_{\langle ij\rangle} ( r_{ij} - b)^2 $, which
acts between adjacent monomers only and ensures chain connectivity.
iv) The external electric potential, $ U_{\mathrm{ext}} / \kB T =
-\sum_i ( q_i \e / \kB T ) \, \vec E \cdot \vec r_i $, with the
electric field directed either parallel ($\vec E_{||}$) or
perpendicular ($\vec E_{\perp}$) to the PE axis.  Periodic boundary
conditions along the PE axis are implemented by a one-dimensional
resummation of the Coulomb interactions~\cite{naji-netz:2006}; the
lateral and all hydrodynamic interactions are treated using the
minimum image convention. Consequential finite-size effects are
discussed in the supplementary information~\cite{sup}.  The PE
electrophoretic mobility $ \mu_{\mathrm{m}} =\ave{ v_{\mathrm{m}}} /
E$ follows from the average monomer velocity along the E-field
direction.  In the absence of curvature, inter-chain and end effects
(\emph{i.e.}\ for high enough salt concentrations) and if orientation
effects are negligible (\emph{i.e.}\ for small E-fields),
$\mu_{\mathrm{m}}$ follows from the parallel and perpendicular
mobilities as $\mu_{\mathrm{m}} = (\mu_{\mathrm{m}}^{||}+2
\mu_{\mathrm{m}}^{\perp})/3$.  In the simulations we accordingly
determine $\mu_{\mathrm{m}}^{||}$ and $\mu_{\mathrm{m}}^{\perp}$
separately by applying E-fields parallel and perpendicular to the PE
axis and measuring the corresponding velocities. Possible non-linear
effects have been carefully checked~\cite{sup}.  The ionic strength
includes contributions from the neutralizing counterions and is
defined as $I = (c_{\mathrm{nct}} q_{\mathrm{ct}}^2 + c_{\mathrm{s}}
q_{\mathrm{ct}}^2+ c_{\mathrm{s}} q_{\mathrm{co}}^2)/2$.

In order to model DNA in aqueous NaCl solution at
$\unit{20}{\celsius}$ we use Stokes radii of Na\textsuperscript{+} and
Cl\textsuperscript{-} as $a_{\mathrm{ct}} = \unit{1.84}{\angstrom}$
and $a_{\mathrm{co}} = \unit{1.29}{\angstrom}$ as obtained from
limiting conductivities~\cite{marcus:1997}, an estimate of
$a_{\mathrm{m}} = \unit{10.47}{\angstrom}$ for the DNA radius and
valencies $q_{\mathrm{ct}} =1$, $q_{\mathrm{co}} = -1$ and
$q_{\mathrm{m}} = -6$. The choice of monomer separation $b =
a_{\mathrm{m}}$ ensures a linear charge density of $q_{\mathrm{m}}/b
\simeq \unit{0.57} { \angstrom^{-1}}$.  Although no bending rigidity
is present in the model, the segment is sufficiently straight due to
electrostatic repulsions, as appropriate for DNA (cf.\
Fig.~\ref{fig:1}).  The simulation cell comprises $10$ DNA monomers,
$60$ neutralizing counterions and $24$ salt pairs. The ionic strength
is varied over the range $I = \unit{19-468}{\milli\mathrm{M}}$ by
adjusting the cell width $D$.  The field strengths applied are $E_{||}
= \unit{27.5 \times 10^6}{\volt \per \metre}$ and $E_{\perp} =
\unit{5.5 \times 10^6}{\volt \per \metre} $.  We use $\epsilon = 5$
for the LJ strength, $K/a_{\mathrm{ct}}^2 = 100$ for the bond
stiffness, and $\eta = \unit{1.003 \times 10^{-3}}{\pascal \,
  \second}$ for the viscosity of water.  The Langevin time-step is
$\unit{0.06-0.12}{\pico\second}$ and simulations are typically run for
$\unit{0.3-4.1}{\micro\second}$.

In Fig.~\ref{fig:2}a we plot the DNA electrophoretic mobility
$\mu_{\mathrm{m}} $ as a function of the ionic strength, $I$, together
with experimental data for long DNA from
Refs.~\cite{hartford-flygare:1975,hoagland-arvanitidou-welch:1999}.
Noting that there are no free fitting parameters and given the
substantial scatter in the experimental data, we conclude that our
coarse-grained DNA model is quite accurate.  The mobility
$\mu_{\mathrm{m}}$ decreases with increasing $I$ which will be
rationalized in terms of hydrodynamic screening effects below.  We
additionally show theoretical results from
Stigter~\cite{schellman-stigter:1977} and Manning~\cite{manning:1981}.

Theoretically, only little attention has been paid to E-field-induced
counterion dynamics in PE solutions. In this context the phenomenon of
counterion condensation at highly charged PEs that are characterized
by a Manning parameter $\xi_{\mathrm{M}} = |q_{\mathrm{ct}}
q_{\mathrm{m}} \lB / b| > 1$ has to be taken into account. For highly
charged PEs such as DNA ($\xi_{\mathrm{M}} = 4.17$) electrostatic
attraction of counterions towards the PE overcomes entropic repulsion
giving rise to increased accumulation of counterions in the very
vicinity of the PE~\cite{manning:1969a,naji-netz:2006}.  In
particular, the assumption that condensed counterions stick to the
PE~\cite{schellman-stigter:1977,manning:1981} has not been
scrutinized, despite experimental evidence that condensed counterions
are not immobilized on the PE surface~\cite{nagasawa-etal:1972}.  In
Fig.~\ref{fig:2}b we show the electrophoretic mobility of two
counterion ensembles, first condensed counterions within a distance
$r^{\ast}_\perp =a_{\mathrm{m}}+4a_{\mathrm{ct}}=
\unit{17.8}{\angstrom}$ from the DNA axis ($\mu_{\mathrm{cct}}$) and
secondly the set of counterions closest to the DNA axis that
neutralize the DNA charge ($\mu_{\mathrm{nct}}$).  At low ionic
strength the hydrodynamic drag exerted by the DNA on the counterions
exceeds the external electric force and the mobility for both sets is
negative, \emph{i.e.}\ the counterions are dragged along by the PE.
At high ionic strength the hydrodynamic interactions are sufficiently
screened so that the electric field dominates and the counterions move
opposite to the DNA.  In fact, a salt and PE charge density dependent
sign reversal of the electrophoretic counterion mobility has been
inferred from transference experiments some time
ago~\cite{nagasawa-etal:1972}.  Direct measurements of counterion
electrophoretic mobilities can in principle be performed with pulsed
field gradient NMR~\cite{boehme-scheler:2004}.

We define the excess contribution of the counterions to the
conductivity of a PE solution as
\begin{equation}
  \label{eq:anct}
  A_{\mathrm{nct}} = %
  (\sigma -q_{\mathrm{m}} \e c_{\mathrm{m}} \mu_{\mathrm{m}} %
  -\sigma_{\mathrm{s}}^0) / c_{\mathrm{nct}}
\end{equation}
where $\sigma$ and $\sigma_{\mathrm{s}}^0$ denote the specific
conductivities of the salt solution with and without the PE chain,
respectively.  In our simulations, $\sigma$ results from the separate
electrophoretic contributions as $\sigma /\e = q_{\mathrm{ct}}
c_{\mathrm{nct}} \mu_{\mathrm{nct}} - q_{\mathrm{m}} c_{\mathrm{m}}
\mu_{\mathrm{m}} + q_{\mathrm{ct}} c_{\mathrm{s}} \mu_{\mathrm{ct}} -
q_{\mathrm{co}} c_{\mathrm{s}} \mu_{\mathrm{co}} $, while the pure
electrolyte conductivity $\sigma_0$ is obtained from separate
simulations as $\sigma_{\mathrm{s}}^0 = q_{\mathrm{ct}} c_{\mathrm{s}}
\mu_{\mathrm{ct}}^0 - q_{\mathrm{co}} c_{\mathrm{s}}
\mu_{\mathrm{co}}^0 $.  As seen in Fig.~\ref{fig:2}c, the counterion
excess conductivity $ A_{\mathrm{nct}}$ increases with increasing salt
concentration and changes sign, and thus directly reflects the
salt-dependent counterion mobility anomaly for the experimentally
easily accessible conductivity.

To gain further insight, we now shift to weakly charged PEs (Manning
parameter $\xi_{\mathrm{M}} < 1$), where the ion distribution around a
PE is correctly described by linear Debye-H{\"u}ckel (DH) theory and
the electrophoretic mobilities of PE and ions can be constructed using
Green's functions~\cite{barrat-joanny:1996}.  The DH ionic charge
distribution around a sphere of radius $a$ and surface charge density
$q \e / 4 \pi a^2$ is for $\kappa a < 1$ given by $ n (r) = - q \e
\kappa^2 \mathrm{e}^{-\kappa (r - a)} / [ 4 \pi r (1 + \kappa a) ]$,
where $\kappa^{-1} = (8\pi \lB I)^{-1/2}$ is the DH screening
length. On the Stokes level, the solvent flow field $\vec u$ induced
by an external electric field $\vec E$ acting on the ionic charge
distribution $n (r)$ is $\vec u(\vec r) = \vec G(\vec r) \, q \e \vec
E$, where the screened hydrodynamic Green's function $G_{\alpha \beta}
(\vec r)$ reads~\cite{sup}
\begin{equation}
  \label{eq:tg}
  G_{\alpha \beta} (\vec r) = %
  A  \bigg( %
  \delta_{\alpha \beta} -3\frac{x_\alpha x_\beta}{r^2} %
  \bigg) %
  + 2 B \frac{x_\alpha x_\beta}{r^2}, 
\end{equation}
where $\alpha, \beta=x,y,z$ and $B = \e^{-\kappa(r-a)} / [ 4\pi \eta r
(1+\kappa a) ]$ and $A = B \, [ 1+\kappa r + \kappa^2 r^2
-\e^{\kappa(r-a)} (1+\kappa a + \kappa^2 a^2 / 3 ) ] / (\kappa^2 r^2
)$.  In the limit of zero salt $\kappa \to 0$, the Stokes solution for
a translating sphere is recovered~\cite{kim-karrila:2005}. For
vanishing radius $a \to 0$, Eq.~(\ref{eq:tg}) reduces to a previously
derived expression~\cite{long-ajdari:2001}.  Noting that
Eq.~(\ref{eq:tg}) fulfills the no-slip condition on the sphere's
surface, its electrophoretic mobility follows from $\mu_{\mathrm{s}}=
q \e \, G_{xx}(r=a)$ as
\begin{equation}
  \label{eq:3}
  \frac{\mu_{\mathrm{s}}}{q \e \mu_0} = %
  \frac{1}{1 +\kappa a} \,,
\end{equation}
which is the classical result derived by Debye and
H{\"u}ckel~\cite{kim-karrila:2005}. Here $\mu_0 = 1/ 6\pi \eta a$ is
the Stokes mobility.

\begin{figure*}
  \centering
  \includegraphics{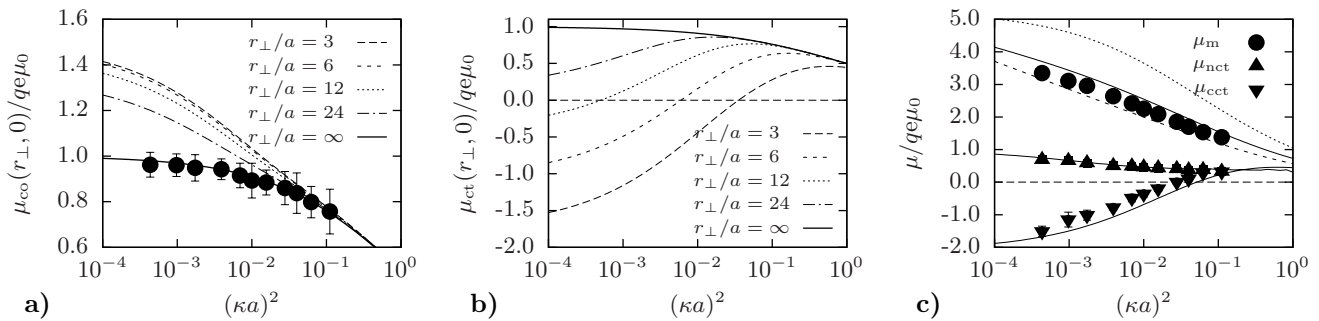}
  \caption{%
    Results for a weakly charged PE with Manning parameter
    $\xi_{\mathrm{M}} = |q_{\mathrm{ct}} q_{\mathrm{m}} \lB / b|=0.4$. %
    a) Coion mobility $\mu_{\mathrm{co}} (r_\perp,0)$ for various
    fixed distances $r_\perp$ from the PE as a function of $(\kappa
    a)^2$. The hydrodynamic drag exerted by the PE on the coions
    increases their mobility as compared to the case without PE,
    Eq.~(\ref{eq:3}) ($r_\perp/a \to \infty$, solid line).  The latter
    compares well with hydrodynamic simulations of a simple salt
    solution (filled symbols). %
    b) Counterion mobility $\mu_{\mathrm{ct}} (r_\perp,0)$ at distance
    $r_\perp$ from the PE which exhibits a sign change. For small
    $r_\perp$ and low salt, counterions are dragged along with the PE
    ($\mu_{\mathrm{ct}} (r_\perp,0) < 0$). %
    c) Comparison of theoretical predictions (solid lines) and
    hydrodynamic simulations (filled symbols) for the PE mobility
    $\mu_{\mathrm{m}}$, neutralizing counterion mobility
    $\mu_{\mathrm{nct}}$ and condensed counterion mobility
    $\mu_{\mathrm{cct}}$. The condensed counterion mobility changes
    sign. We also show the parallel and perpendicular PE mobilities
    $\mu_{\mathrm{m}}^\parallel$ (dotted line) and
    $\mu_{\mathrm{m}}^\perp$ (short dashed line).}
  \label{fig:3}
\end{figure*}

To leading order in $a/r$ the electrophoretic coupling matrix
$\mu_{\mathrm{ss}}^{\alpha \beta} (\vec r)$ ($\alpha, \beta = x, y,
z$) between two charged spheres is obtained via a multipole
expansion~\cite{kim-karrila:2005} as
\begin{equation}
  \label{eq:4}
  \mu_{\mathrm{ss}}^{\alpha \beta} (\vec r) = %
  q \e \left(1 + \frac{a^2}{6} \, \nabla^2_{\vec r} \right) %
  G_{\alpha \beta} (\vec r ) \,.
\end{equation}
Approximating the PE as a straight chain of charged spheres at spacing
$b$ oriented along the $z$-axis, the PE mobilities follow by
superposition as $ \mu_{\mathrm{m}}^\perp = \mu_{\mathrm{s}} + 2
\sum_{j=1}^{\infty} \mu_{\mathrm{ss}}^{xx} (r_\perp=0, b j)$ and $
\mu_{\mathrm{m}}^\parallel = \mu_{\mathrm{s}} + 2 \sum_{j=1}^{\infty}
\mu_{\mathrm{ss}}^{zz} ( r_\perp=0, b j)$. Here $r_\perp$ denotes the
lateral distance from the PE axis.  For the orientationally averaged
mobility $\mu_{\mathrm{m}} = (\mu_{\mathrm{m}}^\parallel +
2\mu_{\mathrm{m}}^\perp)/3$, we obtain the closed-form expression
\begin{equation}
  \label{eq:8}
  \frac{\mu_{\mathrm{m}}}{q \e \mu_0} = %
  \frac{\mu_{\mathrm{s}}}{q \e \mu_0} %
  - \frac{a (6 + \kappa^2 a^2)}{3 b (1 + \kappa a)} \, %
  \e^{\kappa a} \, \ln \left( 1 - \e^{-\kappa b} \right) \,.
\end{equation}
In the limit of low screening, $\kappa a \to 0$, Eq.~\eqref{eq:8}
decays logarithmically with increasing ionic strength as
$\mu_{\mathrm{m}} / q \e \mu_0 = -2 (a / b) \ln (\kappa b)$, in accord
with previous results for weakly charged
PEs~\cite{barrat-joanny:1996,manning:1981}.  In the same fashion the
perpendicular and parallel distance-dependent ion mobilities follow as
$ \mu_{\mathrm{co/ct}}^\perp(r_\perp,z) = \mu_{\mathrm{s}} \pm
\sum_{j=-\infty}^{+ \infty} \mu_{\mathrm{ss}}^{xx} (r_\perp,z + b j) $
and $ \mu_{\mathrm{co/ct}}^\parallel(r_\perp,z) = \mu_{\mathrm{s}} \pm
\sum_{j=-\infty}^{+ \infty} \mu_{\mathrm{ss}}^{zz} (r_\perp,z + b j)
$, respectively; the plus/minus sign applies to coions/counterions.

In Fig.~\ref{fig:3} we compare the foregoing theoretically predicted
electrophoretic mobilities of monomers and ions (obtained by summing
over contributions from $23$ spheres) to the hydrodynamic simulations
of a weakly charged PE with Manning parameter $\xi_{\mathrm{M}} = 0.4$
in a field of strength $aq\e E/\kB T= 0.2$.  The simulation cell
comprises 24 PE monomers, 24 neutralizing counterions and 24 salt
pairs with equal radii $a_{\mathrm{m}} =
a_{\mathrm{co}}=a_{\mathrm{ct}} \equiv a$, valencies $q_{\mathrm{m}} =
q_{\mathrm{co}} = -q_{\mathrm{ct}} \equiv q$ and monomer spacing
$b=2a$.  For intrinsically flexible PEs, the straight PE conformation
in our simulations and theory is realistic only for low enough salt
concentration as long as the effective persistence length is larger
than the screening length. In Fig.~\ref{fig:3}a,b we show the
orientationally averaged coion and counterion mobilities $
\mu_{\mathrm{co/ct}} (r_\perp,0) = [2 \mu_{\mathrm{co/ct}}^\perp
(r_\perp,0) + \mu_{\mathrm{co/ct}}^\parallel (r_\perp,0)]/3$ for fixed
vertical coordinate $z=0$ and various fixed distances $r_\perp $ from
the PE as a function of the rescaled salt concentration $(\kappa a )^2
\sim c_{\mathrm{s}}$. The mobilities of coions are increased and those
of counterions are decreased by the presence of the PE. This
entraining effect is larger for smaller salt concentration and smaller
$r_\perp$. The ion mobilities for $r_\perp= \infty$ reflect pure
electrolyte friction effects and in Fig.~\ref{fig:3}a compare very
well with simulation results for a simple salt solution.  In
Fig.~\ref{fig:3}c we compare analytical predictions for the PE
mobility $\mu_{\mathrm{m}}$, the neutralizing counterion mobility
$\mu_{\mathrm{nct}}$, and the condensed counterion mobility
$\mu_{\mathrm{cct}}$ (obtained from counterions within a shell of
$r_\perp^{\ast} = 5a$ around the PE) with the simulations.  Here
$\mu_{\mathrm{nct}}$ and $\mu_{\mathrm{cct}}$ are obtained from
$\mu_{\mathrm{ct}} (r_{\perp},0)$ by spatially averaging over the DH
counterion distribution around a straight chain of charged spheres at
fixed vertical coordinate $z=0$.  With increasing salt concentration,
$\mu_{\mathrm{cct}}$ changes its sign, similar to the DNA results
(cf.\ Fig.~\ref{fig:2}b). This shows that the salt-induced counterion
mobility anomaly is not restricted to the non-linear regime and is
fully explained by screening effects of the hydrodynamic coupling
tensor.

Our simulation method neglects local solvation and DNA structural
effects.  The good agreement between experimental and simulation
results could imply that those effects are of minor importance for the
electrokinetic behavior.  Nevertheless, an extension of the model to
more realistic charge distributions is planned.  Likewise, the
analytic Green's function approach will be generalized to include
relaxation effects in addition to retardation.

\begin{acknowledgments}
  This work was supported by the German Excellence Initiative via the
  Nanosystems Initiative Munich (NIM).
\end{acknowledgments}


\end{document}